\documentclass[aps,eqsecnum,preprint]{revtex4}
\usepackage{amsmath2000,amssymb,amsfonts}

\newcommand{\di}{\partial}
\newcommand{\benonumber}{\begin{displaymath}}
\newcommand{\eenonumber}{\end{displaymath}}
\newcommand{\be}{\begin{equation}}
\newcommand{\ee}{\end{equation}}
\newcommand{\eref}[1]{(\ref{#1})}
\newcommand{\grad}{\nabla}
\newcommand{\half}{\frac{1}{2}}
\newcommand{\pd}{\ensuremath{\dot{\varphi}}}
\newcommand{\bd}{\ensuremath{\dot{\beta}}}
\newcommand{\sd}{\ensuremath{\dot{\sigma}}}
\newcommand{\Fd}{\ensuremath{\dot{F}}}
\newcommand{\Gd}{\ensuremath{\dot{G}}}
\newcommand{\qd}{\ensuremath{\dot{q}}}
\newcommand{\pdd}{\ensuremath{\ddot{\varphi}}}
\newcommand{\Fdd}{\ensuremath{\ddot{F}}}
\newcommand{\qdd}{\ensuremath{\ddot{q}}}
\newcommand{\pp}{\ensuremath{\varphi'}}
\newcommand{\bp}{\ensuremath{\beta'}}
\renewcommand{\sp}{\ensuremath{\sigma'}}
\newcommand{\Fp}{\ensuremath{F'}}
\newcommand{\Gp}{\ensuremath{G'}}
\newcommand{\qp}{\ensuremath{q'}}

\newcommand{\gp}{\ensuremath{g'}}
\newcommand{\fpp}{\ensuremath{f''}}
\newcommand{\gpp}{\ensuremath{g''}}
\newcommand{\qpp}{\ensuremath{q''}}

\begin{document}
\title{Inhomogeneous M-Theory Cosmologies}
\author                 {Alan A. Coley}
\email                  {aac@mscs.dal.ca}
\affiliation            {Department of Mathematics and Statistics,
                        Dalhousie University, Halifax, N.S.,B3H 3J5, Canada}

\author                 {R. J.  van den Hoogen}
\email                  {rvandenh@stfx.ca}
\affiliation            {Department of Mathematics, Statistics, and Computer
Science,
                        Saint Francis Xavier University, Antigonish, N.S.,
B2G 2W5, Canada}

\begin{abstract}

We study a class of inhomogeneous
and anisotropic $G_2$ string cosmological models.
In the case of separable $G_2$ models we  show that the governing
equations reduce to a system of ordinary differential equations.
We focus on a class of separable $G_2$  M-theory
cosmological models, and study their qualitative behaviour
(a class of models with time-reversed dynamics is also possible).
We find that generically these inhomogeneous M-theory
cosmologies evolve from a spatially inhomogeneous and
negatively curved model with a non-trivial form field
towards spatially flat and  spatially homogeneous dilaton-moduli-vacuum 
solutions with trivial form--fields. 
The late time behaviour is the same as that of spatially homogeneous models previously studied.
However, the inhomogeneities are not dynamically
insignificant at early times in these models.

\end{abstract}
\maketitle{}


\section{String Cosmology}

Non--perturbative  M--theory encompasses and unifies  all  
five anomaly--free, perturbative superstring theories \cite{gsw}
and corresponds to eleven--dimensional supergravity in the
low--energy limit \cite{Witten95}.
In particular, the compactification of M--theory on a circle, $S^1$, 
leads to the type IIA superstring. 
A study of the qualitative cosmological  effects
that can arise in 
M--theory is therefore of considerable importance.
To lowest order (in the inverse string tension), the tree-level
effective action for massless fields contains a dilaton, a
form field (which in $4$-dimensions is dynamically dual to
pseudoscalar axion field)
and (a) stringy cosmological constant(s).  Even in this approximation
the one--loop string equations of motion
for inhomogeneous backgrounds are very difficult to solve, and it is a
useful first
step to consider models in which the homogeneity is broken only in one
spatial direction.
Metrics that
admit two commuting (orthogonally--transitive) space-like Killing vectors
are referred to as $G_2$ spacetimes.

String models admitting an abelian group, $G_2$,  of isometries
have a number of important physical applications. The
spatially homogeneous Bianchi types I--${\rm VII}_h$ and
locally rotationally symmetric (LRS)
types VIII and IX admit a $G_2$ group of isometries
\cite{Tomita78} and so the $G_2$ cosmologies can  be
considered as inhomogeneous generalizations of these Bianchi models.
Non--linear inhomogeneities in the dilaton and axion fields can be
investigated and, in principle, this allows density perturbations
in string--inspired inflationary models such as the pre--big bang
scenario to be studied \cite{Veneziano97,MahOnoVen98}. Given the
potential importance of this scenario it is important to study its
generality with respect to inhomogeneities as well with respect
to anisotropies. The general effects of small inhomogeneities
and anisotropies have been studied by Veneziano \cite{Veneziano97}.

In general relativity (GR) the generic singularity is neither spatially homogeneous or isotropic.
Hence it is of interest to study more general models.
In particular, it has been conjectured
that $G_2$ metrics represent a first approximation to
the general solution of Einstein gravity in the vicinity
of a curvature  singularity \cite{BelKha69,BelKha70,BelKha71,BelLifKha82}.
The high curvature regime
is precisely the regime where stringy deviations from
GR are expected to be significant. The $G_2$ models studied here may
therefore
provide insight into the generic behaviour of cosmologies at very early
times.

A number of exact inhomogeneous
and anisotropic $G_2$ string cosmologies have been found.
Barrow and Kunze studied an inhomogeneous generalization of the
Bianchi type I string cosmology \cite{BarKun97a} and Feinstein, Lazkoz and
Vazquez--Mozo derived a closed, inhomogeneous model by applying duality
transformations on the LRS Bianchi type IX cosmology \cite{FeiLazVaz97}.
Clancy {\em et al.} have found inhomogeneous generalizations
of the Bianchi type ${\rm VI}_h$ universe and have
studied their asymptotic behaviour \cite{ClaFeiLid99b}.

In general, the field equations
reduce to a system of coupled, partial differential
equations in two variables when spatial homogeneity is broken along
a single direction. Unfortunately,
these equations are still very complicated.
However, solutions can be found due to the
non--compact global symmetries of the string effective action.
When the metric admits two commuting space-like Killing vectors,
there exists an infinite--dimensional symmetry on the space of solutions
that may be identified infinitesimally
with the ${\rm O}(2,2)$ current algebra
\cite{Bakas94,Maharana95,Kehagias95}. This
symmetry reduces to the Geroch group,
corresponding to the ${\rm SL}(2,R)$ current
algebra,  when the dilaton and two--form potential
are trivial \cite{Geroch72}, and includes
the global ${\rm SL}(2,R)$ S--duality of
the action.

New inhomogeneous
$G_2$ string cosmologies containing a non--trivial two--form
potential may be generated by an application of both the S-- and T--duality
symmetries
from simpler (dilaton--vacuum) seed solutions.
Lidsey et al. \cite{lidseyetal} discuss the
non--compact, global symmetries of the string effective action in
a variety of settings
and review various methods for solving the Einstein--scalar
field equations utilizing
generating techniques (from solutions with a minimally coupled,
massless scalar field from a vacuum, $G_2$ cosmology).
In particular, Feinstein, Lazkoz and V\'azquez--Mozo \cite{FeiLazVaz97}
present an
algorithm which permits the construction of inhomogeneous string solutions
by employing a Buscher transformation,
inverse scattering techniques
\cite{BelKha70,BelKha71}, followed by the generating technique of
Wainwright, Ince and Marshman \cite{WaiIncMar79}.
Feinstein {\em et al.} employ this algorithm
to generate a closed, inhomogeneous
string cosmology with $S^3$ topology from a LRS Bianchi type IX
solution \cite{FeiLazVaz97,CarChaFei83}. However,
this algorithm involves a number of non--trivial operations, and an
alternative and
more straightforward approach is to apply an ${\rm O}(2,2)$
transformation directly to the seed cosmology \cite{lidseyetal}.

In this paper we shall consider a class of separable string cosmological models
whose governing equations reduce to ordinary differential equations (ODE)
which can be studied by qualitative methods. In particular, we shall focus on a class of M-theory
cosmological models.


\section{String Action}
We consider the general string action in the form \cite{BCL}
\be
\label{action}
S=\int d^4 x \sqrt{-g} \left\{e^{-\Phi} \left[
R +(\grad \Phi)^2 -6 (\grad \beta)^2
-\frac{1}{2} e^{2 \Phi} (\grad \sigma)^2 -2\Lambda
\right] - \half Q^2e^{-6\beta} -\Lambda_{\rm M}\right\},
\ee
in terms of the pseudo--scalar axion field, $\sigma$, the 4-D dilaton field $\Phi$, and the modulus field $\beta$,
where $\Lambda$ and $\Lambda_M$ represent  
cosmological constant terms 
and $Q^2$ may be interpreted as a $0$--form field 
strength.
This is a phenomenological action representing the 
bosonic sector of the effective supergravity action for the
low--energy limit of M--theory and encompasses other string theories \cite{BCL}.
We are particularly interested in the 
class of four--dimensional cosmologies derived from 
the type IIA 
string and M--theory effective actions and which include a non--trivial 
Ramond--Ramond (RR) sector \cite{MT}. In these models a specific compactification from 
eleven to four dimensions
was considered, 
where the topology of the internal dimensions was assumed to be 
a product space consisting of a circle and an isotropic six--torus \cite{MT}; 
this is dynamically equivalent  to compactifications on 
a Calabi--Yau three--fold \cite{kko}. 
The FRW models in this class of cosmologies was studied in \cite{MT}

Defining
\begin{equation}
{\cal L}_{\rm M} = -3 e^{-\Phi} (\grad \beta)^2
-\frac{1}{4} e^{\Phi} (\grad \sigma)^2 -\Lambda e^{-\Phi}
- \frac14 Q^2e^{-6\beta} -\half\Lambda_{\rm M};\qquad
T_{\alpha\beta}  \equiv  g_{\alpha\beta}{\cal L}_{\rm M}
-2 \frac{\di{\cal L}_{\rm M}}{\di g^{\alpha\beta}}, \nonumber\\
\end{equation}
the Euler-Lagrange
equations then lead to the field equations (FE)  \cite{Bthesis}
\begin{subequations}
\label{EFE}
\begin{eqnarray}
G_{\mu\nu} &=& -\grad_\mu\grad_\nu\Phi
+6 \grad_\mu\beta \grad_\nu\beta
+\frac{1}{2} e^{2\Phi} \grad_\mu\sigma\grad_\nu \sigma
\nonumber \\ &&
-\frac12g_{\mu\nu}\Biggl[(\grad\Phi)^2+6  (\grad \beta)^2
+\frac{1}{2} e^{2\Phi} (\grad \sigma)^2 +2\Lambda
+\frac12 Q^2e^{\Phi-6\beta} \nonumber\\
&& \qquad\qquad\qquad+\Lambda_{\rm M}e^{\Phi} -2\Box\Phi\Biggr],\label{Gab}\\
\label{bPhi}
\Box\Phi&=&\half(\grad\Phi)^2 +3 (\grad \beta)^2
-\frac{1}{4} e^{2 \Phi} (\grad \sigma)^2 +\Lambda -\half R, \\
\label{bbeta}
\Box\beta&=&\grad_\mu\Phi\grad^\mu\beta -\frac14 Q^2 e^{\Phi-6\beta}, \\
\label{bsigma}
\Box\sigma&=&-\grad_\mu\Phi\grad^\mu\sigma.
\end{eqnarray}
\end{subequations}

In the above, greek indices take on values $0,1,2,3$, and units are chosen so that $16\pi \hat G=1$.


\section{$G_2$ Cosmologies}

Let us examine \eref{EFE} within the context of $G_2$ cosmological models
described by the line element
\be
ds^2= e^{2F}\left(-dt^2+dz^2\right)+e^G\left(e^q dx^2+e^{-q}dy^2\right),
\label{G2}
\ee
where the metric functions $\{F,G,q\}$ and the string functions
$\{\Phi, \beta, \sigma\}$ are all functions of $t$ and $z$ only.
For any $q(t,z)$, we define
$\displaystyle \dot{q} \equiv \frac{\partial q}
{\partial t}, \enskip q^\prime \equiv \frac{\partial q}
{\partial z}$, and
$\Delta^2 q \equiv   \qdd - \qpp.$
The local behaviour
of these models is determined by the gradient $B_{\mu}
\equiv \partial_{\mu} G$, and
cosmological solutions arise if $B_{\mu}$ is globally time-like.

Also,  the Ricci scalar is given by
\be
R= \half e^{-2F}\left[4\Delta^2 G+4\Delta^2 F+3\left(\Gd^2-\Gp^2 \right)
+\left(\qd^2-\qp^2 \right)\right].
\ee
Using these expressions,
and defining the  modified dilaton field,
\be
\varphi\equiv \Phi-F-G,
\ee
the field equations become
\begin{subequations}
\label{FE}
\begin{eqnarray}
\nonumber
\Delta^2\varphi &=& \frac12\left[\left(\pd+\Fd\right)^2-
\left(\pp+\Fp\right)^2 \right]
                    +\frac14 \left(\Gd^2-\Gp^2\right) +\frac14
\left(\qd^2-\qp^2\right)\\
                &&  +3 \left(\bd^2-\bp^2\right) -\frac14e^{2\varphi+2F+2G}
\left(\sd^2-\sp^2\right) -\Lambda e^{2F},\\
\nonumber \\
\Delta^2\beta&=& \left[\left(\pd+\Fd\right)\bd- \left(\pp+\Fp\right)\bp
\right] +\frac14 Q^2e^{\varphi+3F+G-6\beta}, \\
\nonumber \\
\Delta^2\sigma&=& -\left[\left(\pd+\Fd+2\Gd\right)\sd-
\left(\pp+\Fp+2\Gp\right)\sp \right], \\
\nonumber \\
\Delta^2 q &=& \left[\left(\pd+\Fd\right)\qd- \left(\pp+\Fp\right)\qp
\right], \\
\nonumber \\
\Delta^2 F  &=& \half\left(\pd+\Fd\right)^2-\half\left(\pp+\Fp\right)^2
                -\frac14 \left(\Gd^2-\Gp^2\right) -\frac14
\left(\qd^2-\qp^2\right)\nonumber \\
\nonumber   &&  -3 \left(\bd^2-\bp^2\right) +\frac14e^{2\varphi+2F+2G}
\left(\sd^2-\sp^2\right)
                -\Lambda e^{2F}\\
            &&  -\frac12Q^2 e^{\varphi+3F+G-6\beta}-\Lambda_{\rm M}
e^{\varphi
            +3F +G},\\
\nonumber \\
\nonumber \Delta^2 G  &=& \left[\left(\pd+\Fd\right)\Gd -
\left(\pp+\Fp\right)\Gp\right] + \left(\sd^2-\sp^2\right)
                e^{2\varphi+2F+2G} \\
            &&    -\frac12Q^2 e^{\varphi+3F+G-6\beta}-\Lambda_{\rm M}
e^{\varphi +3F +G}, \\
\nonumber \\
\nonumber \left(\pd+\Fd\right)^{.}&+&\left(\pp+\Fp\right)' = 2
\left(\pd+\Fd\right)\Fd
                +2 \left(\pp+\Fp\right)\Fp +\half\left(\Gd^2+\Gp^2\right)\\
             && +\frac12\left(\qd^2+\qp^2\right) +6\left(\bd^2+\bp^2 \right)
                +\frac12\left(\sd^2+\sp^2\right)e^{2\varphi+2F+2G},
\label{F2}\\
\nonumber \\
\nonumber \left(\pd+\Fd\right)' &=&
\frac12\Gd\Gp+\frac12\qd\qp+6\dot\beta\beta'
                +\Fp\left(\pd+\Fd\right) +\Fd\left(\pp+\Fp\right)\\
            && +\frac{1}{2}e^{2\varphi+2F +2G}\sd\sp.
\end{eqnarray}
\end{subequations}

These equations reduce to those in \cite{BCL,MT} in the appropriate limits.


\section{Separable $G_2$ String Cosmologies}

\subsection {General Case}

Let us assume separability of the metric functions of the form
\begin{eqnarray*}
F(t,z)&\equiv& F(t)+ f(z),\\
G(t,z)&\equiv& G(t) + g(z), \\
q(t,z)&\equiv& q(t)+ \nu(z), \\
\end{eqnarray*}
and appropriate separability conditions on the
matter fields $\Phi(t,z),
\beta(t,z),
\sigma(t,z)$.
Then the Ricci Scalar is given by
$$ R = \frac{1}{2}e^{-2F-2f}[4 \ddot{G} + 4\ddot{F} + 3\dot{G}^2
+\dot{q}^2 - (4\gpp + 4\fpp + 3\gp^2 + {\nu'}^2)]. $$
If
\begin{equation}
 4\gpp + 4\fpp + 3\gp^2 + {\nu'}^2 = C, \label{a}
 \end{equation}
where $C$ is a constant, then we obtain a condition which constrains the
spatial dependence of the metric. The Ricci Scalar is then given by
$$ R = \frac{1}{2}e^{-2F-2f}[4 \ddot{G} + 4\ddot{F} + 3\dot{G}^2 + \dot{q}^2
-C]. $$
Putting this expression for the Ricci Scalar into the action \eref{action},  the spatial dependence 
of the geometrical terms can be eliminated (by integration over the spatial coordinates
in the action).
 After applying any further separability conditions
(on the matter fields), the resulting FE will  be a system of ODEs. Note
that the
effect of the spatial dependence is to add a further contribution ($C$) to the
cosmological constant $\Lambda$ in the action.

\subsection{Specific Example: Linear Dependence in $z$}

In an attempt to remove the $z$-dependence, let us assume separability of
the form
\begin{eqnarray*}
F(t,z)&\equiv& F(t)+\half cz,\\
G(t,z)&\equiv& G(t), \\
q(t,z)&\equiv& q(t)+az, \\
\Phi(t,z)&\equiv& \Phi(t)+mz, \\
\beta(t,z)&\equiv& \beta(t)+nz, \\
\sigma(t,z)&\equiv& \sigma(t)+lz, \\
\end{eqnarray*}
where $a, c, l, m, n$ are constants ($a^2$ is equivalent to the constant $C$ in equation (\ref{a})), and therefore
\benonumber
\varphi(t,z) = \Phi(t)-F(t)-G(t)+\left(m-\frac12c\right)z \equiv
\varphi(t)+\left(m-\frac12c\right)z.
\eenonumber
With the above assumptions, the metric becomes an extension of the inhomogeneous scalar-field $G_2$ solutions found by Feinstein and Ibanez \cite{FeinsteinIbanez93} to M-Theoretical models. In addition, for particular values of the parameters, the metric reduces to spatially homogeneous Bianchi $I$, $III$ and $VI_0$ models.  
Hence,
\begin{subequations}
\begin{eqnarray}
\nonumber
\ddot\varphi &=& \half\left(\pd+\Fd\right)^2
+\frac14 \Gd^2 +\frac14 \qd^2
+3 \bd^2 -\frac14e^{2\varphi+2F+2G+2mz}
\left(\sd^2-l^2\right)\\ &&  -\Lambda e^{2F+cz}-
\frac14\left(a^2+2m^2+12n^2\right),\label{p1}\\
\ddot\beta&=& \left(\pd+\Fd\right)\bd+\frac14 Q^2
e^{\left[\varphi+3F+G-6\beta+\left(c+m-6n\right)z\right]}-mn , \\
\ddot\sigma&=& -\left(\pd+\Fd+2\Gd\right)\sd+ml, \\
\ddot q &=& \left(\pd+\Fd\right)\qd- ma, \\
\nonumber
\ddot F  &=&
\half\left(\pd+\Fd\right)^2
-\frac14 \Gd^2 -\frac14 \qd^2 -3 \bd^2
+\frac14e^{2\varphi+2F+2G+2mz} \left(\sd^2-l\right)\\
\nonumber
&&
-\Lambda e^{2F+cz}-\frac12Q^2 e^{\left[\varphi+3F+G-6\beta+(c+m-6n)z
\right]}
-\Lambda_{\rm M} e^{\varphi +3F +G+(c+m)z}\nonumber\\
&&
+\frac14(a^2-2m^2+12n^2),\label{F1}\\
\nonumber
\ddot G  &=& \left(\pd+\Fd\right)\Gd +\left(\sd^2-l^2\right)
e^{2\varphi+2F+2G+2mz} \nonumber\\
&& -\frac12Q^2 e^{\left[\varphi+3F+G-6\beta +(c+m-6n)z\right]}
-\Lambda_{\rm M} e^{\varphi +3F +G+(c+m)z}, \\
\nonumber
\left(\Fdd+\pdd\right) &=&
2 \left(\Fd+\pd+m\right)\left(\Fd+\half c\right)
+\half\Gd^2\\
\nonumber &&
+\frac12\left(\qd+a\right)^2 +6\left(\bd+n \right)^2
+\frac12\left(\sd+l\right)^2e^{2\varphi+2F+2G+2mz}. \label{pF1}\\
\end{eqnarray}
\end{subequations}

Note that the 
constraint equation can be rewritten as:
\begin{eqnarray}
0&=&\left(\Fd+m\right)^2-\pd^2 +c\left(\Fd+\pd+m\right) +\half\Gd^2
+\frac12\left(\qd+a\right)^2 \nonumber\\
&& +6\left(\bd+n \right)^2
+\frac12\left(\sd+l\right)^2e^{2\varphi+2F+2G+2mz}\nonumber\\
&& +\Lambda e^{2F+cz}+\frac12Q^2
e^{\left[\varphi+3F+G-6\beta+(c+m-6n)z\right]}
+\Lambda_{\rm M} e^{\varphi +3F +G+(c+m)z}.\label{cons1}
\end{eqnarray}

In order for the FE to be independent of $z$, it is
necessary
that $m=0$ and that either $c=0$ or $c=6n$.  Furthermore if $c=0$ then we have
that either $n=0$ or $Q=0$. In the $c=6n$ case, we have that
$\Lambda=\Lambda_{\rm M}=0$.  It is the latter case that is of interest to
us here.  From  here forth we shall assume that $c=6n$ and that
$\Lambda=\Lambda_{\rm M}=0$.  This particular subcase, which is of relevence
in M-theory cosmology, is of special
physical interest. (The resulting FE in the remaining cases are displayed in \cite{Bthesis}).


\section{Inhomogeneous M-Theory Cosmological models}

Substituting $c=6n$, $m=0$, $\Lambda=0$, $\Lambda_M=0$ into \eref{FE}, \eref{cons1} (and
taking the linear combination [\eref{pF1}-\eref{F1}-\eref{p1}]) we obtain
the following system of ODE with two
constraints:
\begin{subequations}
\begin{eqnarray}
\ddot \varphi &=& \frac{1}{4}\Bigl( \dot q^2 + \dot G^2 - a^2+2(\dot \varphi+\dot
F)^2+12\dot \beta^2-12n^2 \nonumber\\
    && \qquad\qquad\qquad\qquad\qquad\qquad +e^{2\varphi + 2F +2G}(l^2-\dot
\sigma^2) \Bigr)\\
\ddot \beta &=& \dot \beta(\dot \varphi+\dot
F)+\frac{1}{4}Q^2e^{-6\beta+\varphi+3F+G}\\
\ddot \sigma &=& -\dot\sigma(\dot \varphi+\dot F+2\dot G)\\
\ddot q &=& \dot q(\dot \varphi + \dot F)\\
\ddot F &=& \frac{1}{4}\Bigl( 3a^2+\dot q^2 +\dot G^2 +36n^2+12\dot \beta^2
+e^{2\varphi+2F+2G}(l^2+3\dot \sigma^2)\nonumber\\
        && \qquad\qquad\qquad\qquad\qquad\qquad-2(\dot \varphi-\dot F)^2+8\dot
F^2 \Bigr)\\
\ddot G &=& \dot G(\dot \varphi+\dot F)-e^{2\varphi + 2F +2G}(l^2-\dot
\sigma^2) -\frac{1}{2}Q^2e^{-6\beta+\varphi+3F+G}\\
0 &=& 2\dot\varphi^2-\dot G^2-\dot q^2-12\dot \beta^2-2\dot
F^2-12n^2 -a^2\nonumber\\
        &&
\qquad\qquad\qquad\qquad\qquad\qquad -e^{2\varphi+2F+2G}(l^2+\dot\sigma^2)
            -Q^2e^{-6\beta+\varphi+3F+G}\label{Q2}\\
0&=&a\dot q + le^{2\varphi+2F+2G}\dot\sigma+6n(\dot \varphi +\dot F+2\dot \beta)
\end{eqnarray}
\end{subequations}

From equation \eref{Q2} we are able solve for and make a global substitution
for the quantity $Q^2e^{-6\beta+\varphi+3F+G}$.   Making this substitution we
have the following system of ODEs:
\begin{subequations}
\label{DS1}
\begin{eqnarray}
\ddot \varphi &=& \frac{1}{4}\left( \dot q^2 + \dot G^2 - a^2+2(\dot \varphi+\dot
F)^2+12\dot \beta^2-12n^2+e^{2\varphi + 2F +2G}(l^2-\dot \sigma^2) \right)\\
\ddot \beta &=& \frac{1}{4}\left( 2\dot\varphi^2-\dot G^2-\dot q^2-12\dot
\beta^2-2\dot F^2-12n^2
        -a^2-e^{2\varphi+2F+2G}(l^2+\dot \sigma^2)\right)\nonumber\\
        &&+ \dot \beta(\dot \varphi+\dot F)\\
\ddot \sigma &=& -\dot\sigma(\dot \varphi+\dot F+2\dot G)\\
\ddot q &=& \dot q(\dot \varphi + \dot F)\\
\ddot F &=& \frac{1}{4}\left( 3a^2+\dot q^2 +\dot G^2 +36n^2+12\dot \beta^2
+e^{2\varphi+2F+2G}(l^2+3\dot \sigma^2)-2(\dot \varphi-\dot F)^2+8\dot F^2
\right)\\
\ddot G &=& \dot G(\dot \varphi+\dot F)-e^{2\varphi + 2F +2G}(l^2-\dot \sigma^2)
\nonumber\\
&&-\frac{1}{2}\left(2\dot\varphi^2-\dot G^2-\dot q^2-12\dot \beta^2-2\dot
F^2-12n^2 -a^2-e^{2\varphi+2F+2G}(l^2+\dot \sigma^2)\right)\\
0&=&a\dot q + le^{2\varphi+2F+2G}\dot\sigma+6n(\dot \varphi +\dot F+2\dot \beta)
\end{eqnarray}
\end{subequations}

From the constraint \eref{Q2} we see that if $\dot \varphi = 0$, then all of the other state variables
must be simultaneously zero, which can only occur at an equilibrium point of the system.
Hence  $\dot \varphi$ must be positive (or negative) throughout the physical phase space.
Here we shall assume $\dot \varphi>0$ (the case $\dot \varphi<0$ can be obtained by a time reversal -- see later).

We define new variables of the form
\begin{eqnarray*}
\tilde F &=& \frac{\dot F}{\dot \varphi}\qquad\qquad
\tilde G = \frac{1}{\sqrt{2}}\frac{\dot G}{\dot \varphi}\qquad\qquad
\tilde q = \frac{1}{\sqrt{2}}\frac{\dot q}{\dot \varphi} \qquad\qquad
\tilde \sigma  =  \frac{1}{\sqrt{2}}e^{\varphi+F+G}\frac{\dot \sigma}{\dot
\varphi}\\
\tilde \beta &=& \sqrt{6}\frac{\dot \beta}{\dot \varphi}\qquad\qquad
\tilde \Psi_1 = \sqrt{\frac{a^2+12n^2}{2}}\frac{1}{\dot\varphi}\qquad\qquad
\tilde \Psi_2 = \frac{1}{\sqrt{2}}e^{\varphi+F+G} \frac{l}{\dot\varphi}
\end{eqnarray*}
and a new time variable
\begin{equation} 
\frac{dt}{d\tau}=\frac{1}{\dot
\varphi}\label{newtime}
\end{equation}

The variables are chosen so that the transformed dynamical system has a
compactified phase space. This property comes from the fact that
$Q^2e^{-6\beta+\varphi+3F+G}\geq 0$ which implies that equation \eref{Q2}
yields
\begin{equation}
1\geq \tilde F^2 + \tilde G^2 +\tilde q^2 +\tilde \sigma^2 +\tilde \beta^2
+\tilde \Psi_1^2 +\tilde \Psi_2^2
\end{equation}

The dynamical system \eref{DS1} becomes
\begin{subequations}
\label{DS2}
\begin{eqnarray}
\frac{d\tilde F}{d\tau}  & = & \tilde F(2\tilde F- \tilde r)+\frac{1}{2}\left(\tilde q^2
+\tilde G^2+\tilde \beta^2+3\tilde\sigma^2
+3\tilde\Psi_1^2+\tilde\Psi_2^2-(1-\tilde F)^2\right),\\
\frac{d\tilde G}{d\tau} & = & \tilde G (1 + \tilde F -\tilde
r) -\frac{1}{\sqrt{2}}\left( 1-\tilde F^2 - \tilde G^2 -\tilde q^2 -3\tilde
\sigma^2 -\tilde \beta^2 -\tilde \Psi_1^2 +\tilde \Psi_2^2\right),\\
\frac{d\tilde q}{d\tau} & = & \tilde q (1 + \tilde F - \tilde r),\\
\frac{d\tilde \sigma}{d\tau} & = & -\tilde\sigma(\sqrt{2}\tilde G + \tilde r),\\
\frac{d\tilde \beta}{d\tau} & = & \tilde \beta(1+\tilde F - \tilde r)
+\frac{\sqrt{6}}{2}\left( 1-\tilde F^2 - \tilde G^2 -\tilde q^2 -\tilde
\sigma^2 -\tilde \beta^2 -\tilde \Psi_1^2 -\tilde \Psi_2^2\right),\\
\frac{d\tilde \Psi_1}{d\tau} & =& -\tilde \Psi_1 \tilde r,\\
\frac{d\tilde \Psi_2}{d\tau} & = & \tilde\Psi_2 (1+\tilde F+\sqrt{2}\tilde G-\tilde r),
\end{eqnarray}
\end{subequations}
and
$$\tilde r = \frac{1}{2}\left[(1+\tilde F)^2+\tilde G^2+\tilde q^2+\tilde
\beta^2+\tilde \Psi_2^2-\tilde \Psi_1^2-\tilde\sigma^2\right],$$
where the constraint equation becomes
\begin{equation}
0 = \sqrt{\frac{2}{a^2+12n^2}}\tilde\Psi_1\left[\sqrt{2}a\tilde q +
6n(1+\tilde F+\frac{2}{\sqrt{6}}\tilde\beta)\right]+2\tilde \Psi_2\tilde\sigma.
\label{constraint}
\end{equation}

There exists a first integral in the physical phase space ($\tilde q \not = 0, \tilde \Psi_1 \not = 0$) for this system.  The function 
$$M=\frac{\tilde\sigma \tilde\Psi_2}{\tilde q\tilde\Psi_1}$$
is constant
, i.e., $M'=0$.  This implies a first integral for the original
system of ordinary differential equations 
(\ref{DS1}) $$\dot q = C \dot\sigma e^{2\varphi+2F+2G}$$ where $C$ is a constant.

\subsubsection{Invariant Sets, Monotonic Functions}

We first recall that the phase space for this dynamical system is the
interior and boundary of the compact set given by
\begin{equation}
1\geq \tilde F^2 + \tilde G^2 +\tilde q^2 +\tilde \sigma^2 +\tilde \beta^2
+\tilde \Psi_1^2 +\tilde \Psi_2^2
\end{equation}
Various hyperplanes divide the phase space into a number of different
regions, they are $\tilde q=0$, $\tilde \sigma =0$, $\tilde \Psi_1=0$ and
$\tilde \Psi_2=0$ hyperplanes.
We note that $\tilde \Psi_1=0$ divides the phase space into two distinct
regions $\tilde \Psi_1<0$ and $\tilde \Psi_1>0$.  The dynamics in the
invariant set $\tilde\Psi_1<0$ is the time reversal of the dynamics in the
invariant set $\tilde\Psi_1>0$ (see \eref{newtime}).

Consider the function $$M_1=\frac{\tilde\sigma^2\tilde\Psi_2^2}{\tilde
q^4}$$ and its derivative $$\frac{dM_1}{d\tau}=-2M_1(1+\tilde F)$$
We easily see that this function is monotonically decreasing in the
invariant set $\tilde q\not = 0, \tilde \sigma\not=0,\tilde\Psi_2\not=0$.
Therefore, we can conclude that there are no closed or periodic orbits in
the seven dimensional phase space, except possibly on the lower dimensional
boundaries of this seven dimensional invariant set.

We restrict ourselves now to the invariant set $\tilde \sigma = 0$.
Consider the function $$M_2=\frac{\tilde\Psi_1^2}{\tilde q^2}$$ and its
derivative $$\frac{dM_2}{d\tau}=-2M_2(1+\tilde F)$$
We easily see that this function is monotonically decreasing in the
invariant set $\tilde q\not = 0,\tilde\Psi_1\not=0$.  Therefore, we can
conclude that there are no closed or periodic orbits in this six dimensional
phase space, except possibly on the lower dimensional boundaries of this six
dimensional invariant set.

In the six dimensional invariant set $\tilde\Psi_2=0$, the function
$$M_3=\frac{\tilde\Psi_1^2}{\tilde q^2}$$ has the derivative $$\frac{d M_3}{d\tau}=-2M_3(1+\tilde F)$$ which is monotonically decreasing in the set $\tilde \Psi_2=0,\tilde q\not = 0, \tilde \Psi_1\not = 0$.  Therefore we conclude that there are no closed or periodic orbits in this six-dimensional invariant set.

In the six dimensional invariant set $\tilde q=0$, the function $$M_4=\frac{\tilde \sigma^2\tilde\Psi_2^2}{\tilde\Psi_1^4}$$ has the derivative $$\frac{d M_4}{d\tau}=2M_4(1+\tilde F)$$ which is monotonically increasing in the set $\tilde q = 0,\tilde \sigma \not = 0, \tilde \Psi_2\not = 0$.  Therefore we can conclude that there are no closed or periodic orbits in this six dimensional set.

With the existence of these monotonic functions $M_1, M_2, M_3$, and $M_4$, we can conclude that there are no closed or periodic orbits in the physical six dimensional phase space, [except possibly on lower dimensional (less than 5) invariant sets].

The zero-curvature spatially homogeneous and isotropic spacetimes are contained in the set $\tilde q = \tilde \Psi_1= \tilde \Psi_2 =\tilde G-\sqrt{2}\tilde F =0$ union $\tilde q = \tilde \Psi_1=\tilde \sigma =\tilde G-\sqrt{2}\tilde F =0$.   

The matter fields in (\ref{action}) satisfy various energy conditions.  For example, the positivity of the kinetic energy of the pseudo-axion scalar field, $\sigma$, demands that $\dot \sigma^2-l^2\geq 0$ (i.e., $\tilde \sigma^2-\tilde\Psi_2^2\geq 0$).  However, we note that $\tilde \sigma^2-\tilde\Psi_2^2=0$ is not an invariant set.

\subsubsection{Equilibrium Points and Exact Solutions}

There are two equilibrium points and one three-dimensional equilibrium set.

${\bullet}$ The three-dimensional equilibrium set is given by
$$ \{\tilde F ^2+\tilde G^2 +\tilde q^2 + \tilde \beta^2=1, \tilde \sigma=0,
\tilde \Psi_1 =0, \tilde\Psi_2 =0 \}$$
Note that since both $\tilde \Psi_1=0$ and $\tilde \Psi_2=0$ we necessarily have
that $a=n=l=0$.  At this point the value of $\tilde r=1+\tilde
F_0$.  The exact solution is then
\begin{eqnarray*}
\varphi(t) & = & h_1-\frac{1}{\tilde r}\ln(\tilde rt+h_0),\\
F(t) & = & F_0(h_1-\frac{1}{\tilde r}\ln(\tilde rt+h_0))+F_1\\
G(t) & = & \sqrt{2}G_0(h_1-\frac{1}{\tilde r}\ln(\tilde rt+h_0))+G_1\\
q(t) & = & \sqrt{2}q_0(h_1-\frac{1}{\tilde r}\ln(\tilde rt+h_0))+q_1\\
\sigma(t) & = & \sigma_1 \\
\beta(t) & = & \frac{1}{\sqrt{6}}\beta_0(h_1-\frac{1}{\tilde r}\ln(\tilde
rt+h_0))+\beta_1
\end{eqnarray*}
where $F_0^2+G_0^2+q_0^2+\beta_0^2=1$ and where
$F_1,G_1,q_1,\sigma_1,\beta_1,h_1$ and $h_0$ are integration constants.

Since $a=n=l=0$, this metric is spatially homogeneous (and flat). Since $\tilde \sigma=0$
and $Q=0$ (which follows from the other conditions), this
equilibrium set represents  spatially flat solutions where the form--fields 
(the axion field and the four--form field strength) are trivial and 
only the dilaton and moduli fields are dynamically important. These solutions are
known as the `dilaton--moduli--vacuum' 
solutions (and their analytical form is given in \cite{coplahwan}).

Recall that the dynamics of these models is restricted by the constraint
given by equation (\ref{constraint}).  At these equilibrium points we are able
to locally solve for the value of $\tilde \Psi_1$ and substitute into the
remaining equations.  The eigenvalues in the six-dimensional constraint
surface are
$$0,0,0,\sqrt{2}\tilde G,-\sqrt{2}\tilde G-1-\tilde F,-1+\tilde F+\sqrt{2}\tilde G-\sqrt{6}\tilde \beta$$
The three zero eigenvalues correspond to the fact that this equilibrium set
is three-dimensional.  It is clear that a subset of this equilibrium set
will act as saddles of varying degree of stability, while another subset 
will act as sinks. Consequently, a
subset of these dilaton-moduli-vacuum 
solutions with trivial form fields are sinks in the physical phase space,
even in the presence of (negative)
spatial curvature, and are thus
generic attracting solutions.

${\bullet}$ A second equilibrium point is
$$\{\tilde F=-\frac{2}{3}, \tilde G=0, \tilde q=0, \tilde
\beta= -\frac{1}{\sqrt{6}},\tilde \sigma=\pm\frac{\sqrt{2}}{6},\tilde
\Psi_1=\frac{\sqrt{2}}{3},\tilde \Psi_2=0\}$$
Note that since $\tilde \Psi_2=0$  we have necessarily that $l=0$.  At this
point the value of $\tilde r=0$.  [Note, since the dynamical system \eref{DS2} is invariant under the transformation $(\tilde \Psi_1,\tilde \Psi_2) \to (-\tilde \Psi_1,-\tilde \Psi_2)$ there exists a corresponding equilibrium point with a $\tilde \Psi_1 =-\sqrt{2}/3$.] The exact solution is then
\begin{eqnarray*}
\varphi(t) & = & h_0t+h_1,\\
F(t) & = & -\frac{2}{3}(h_0t+h_1)+F_1\\
G(t) & = &  G_1\\
q(t) & = &  q_1\\
\sigma(t) & =& \mp e^{-1/3(h_0t+h_1)-(F_1+G_1)}+ \sigma_1 \\
\beta(t) & = & -\frac{1}{6}(h_0t+h_1)+\beta_1
\end{eqnarray*}
where $h_0=\frac{3}{2}\sqrt{a^2+12n^2}$, $Q^2=\frac{1}{2}(a^2+12n^2)e^{6\beta_1-3F_1-G_1}$ and
$F_1,G_1,q_1,\sigma_1,\beta_1,h_1$ are all constants.
In this situation the variable $\tilde q$ can be eliminated.  The eigenvalues
restricted to the constraint surface  are
$$\frac{1}{3},\frac{1}{3},\frac{1}{6}(1\pm\sqrt{15+8\sqrt{2}}i),\frac{1}{6}(
1\pm\sqrt{15-8\sqrt{2}}i)$$
This point represents a past attractor or a source. This corresponds to a
spatially non-vacuum inhomogeneous model with a diagonal Einstein Tensor, having negative curvature.

The line element corresponding to this solution (after a few coordinate
redefinitions) is
\begin{equation}
ds^2= C^2 e^{-2\sqrt{a^2+12n^2}t + 6nz}\left(-dt^2+dz^2\right)+\left(e^{az} dx^2+e^{-az}dy^2\right),
\end{equation}

${\bullet}$ The third equilibrium point is
$$\{\tilde F=-\frac{5}{7}, \tilde G=\frac{\sqrt{2}}{7}, \tilde q=0, \tilde
\beta= -\frac{\sqrt{6}}{7},\tilde \sigma=0,\tilde
\Psi_1=\frac{2\sqrt{3}}{7},\tilde \Psi_2=0\}$$
Note that since $\tilde \Psi_2=0$  we have necessarily that $l=0$.  At this
point the value of $\tilde r=0$.  [Note, since the dynamical system \eref{DS2} is invariant under the transformation $(\tilde \Psi_1,\tilde \Psi_2) \to (-\tilde \Psi_1,-\tilde \Psi_2)$ there exists a corresponding equilibrium point with a $\tilde \Psi_1 =-2\sqrt{3}/7$.]  The exact solution is then
\begin{eqnarray*}
\varphi(t) & = & h_0t+h_1,\\
F(t) & = & -\frac{5}{7}(h_0t+h_1)+F_1\\
G(t) & = &  \frac{2}{7}(h_0t+h_1)+G_1\\
q(t) & = &  q_1\\
\sigma(t) & =& \sigma_1 \\
\beta(t) & = & -\frac{1}{7}(h_0t+h_1)+\beta_1
\end{eqnarray*}
where $h_0=\frac{7}{2\sqrt{3}}\sqrt{a^2+12n^2}$, $Q^2=\frac{1}{3}(a^2+12n^2)e^{6\beta_1-3F_1-G_1}$ and
$F_1,G_1,q_1,\sigma_1,\beta_1,h_1$ are all constants.
This solution is a curved inhomogeneous model with a trivial axion field.
In this situation the variable $\tilde q$ can be eliminated.  The eigenvalues
restricted to the constraint surface  are
$$\frac{2}{7},\frac{2}{7},\frac{4}{7},-\frac{2}{7},\frac{1}{7}(1\pm\sqrt{23}i
)$$
This point is always a saddle.

\section{Discussion}

We have studied several classes of inhomogeneous string models whose governing
equations reduce to ODE. In particular,
we have found that generically solutions of the class of separable $G_2$ inhomogeneous M-theory
cosmologies studied  evolve from a spatially inhomogeneous and
negatively curved model with a non-trivial form fields
towards (a subset) of  spatially flat and  spatially homogeneous dilaton-moduli-vacuum solutions where the form--fields 
(the axion field and the four--form field strength) are trivial and 
only the dilaton and moduli fields are dynamically important.
This late time behaviour is the same as that of the spatially homogeneous models
studied previously. However, in these models the inhomogeneities are not dynamically
insignificant at early times, and the models asymptote (in the past) toward a new class of 
inhomogeneous cosmological models.

As noted earlier, the time--reversed dynamics of the $\dot \varphi>0$
models we have considered thus far 
is equivalent to the dynamics of the case where $\dot \varphi<0$. This follows by 
redefining the time variable according to
 $\frac{dt}{d\tau} = -\frac{1}{\dot \varphi}$
and appropriate definitions of the other state variables.
The evolution equations will have an `overall' change in sign, and hence
the equilibrium points are identical in both cases, 
but the eigenvalues have opposite signs. 
Consequently, the dynamics 
of the $\dot \varphi  < 0$ models is the time reversal of the  
$\dot \varphi> 0$ models and
the time-reversed dynamics of the above 
class of models is deduced by interchanging 
the sources and sinks and reinterpreting 
expanding solutions in terms of contracting ones, and vice--versa.

Although at late times (in the $\dot \varphi>0$ models) the inhomogeneities decay,
the inhomogeneities are important at intermediate times and, in particular, at early
times. Thus the qualitative features of the models are quite different to those of
spatially homogeneous models studied previously. For example, in a study of FRW models
\cite{MT} it was found that
all negatively-curved FRW models evolve from the solution
corresponding to a global source in which the curvature is (negative and) dynamically important
(but with a trivial axion field)
towards the dilaton-moduli-vacuum solutions \cite{coplahwan},
even in the presence of 
spatial curvature.
The physical interpretation of these models, where both the 
NS--NS two--form potential and RR three--form potential 
are dynamically significant, was discussed in \cite{MT}, with particular emphasis
on the fact that the RR field causes the universe to 
collapse, but the NS--NS field has the opposite effect, whereby the interplay 
between these two fields leads to the models undergoing bounces. 
In the models under investigation here, orbits in the full phase space (with $\dot \varphi$ monotone) approach the dilaton-moduli-vacuum solution on the zero-curvature boundary (at late times) and again exhibit a 'bouncing' behaviour;  this bouncing behaviour is the result of the orbits shadowing orbits in the boundary that are constantly being redirected to saddle points of the same or higher stability until it reaches a stable equilibrium.

\begin{acknowledgments}
Both AAC and RJvdH are supported by research grants through Natural Sciences and Engineering Research Council of Canada.  RJvdH wishes to acknowledge the support of the University Council on Research at St. Francis Xavier University.
\end{acknowledgments}

\centerline{{\bf References}}


\begin{thebibliography}{10}

\bibitem{gsw} M. B. Green, J. H. Schwarz, and E. Witten,
{\em Superstring Theory}, in 2 vols., (Cambridge University 
Press, Cambridge, 1987); J. Polchinski, {\em String Theory}, in 2 vols., 
(Cambridge University Press, Cambridge, 1998). 

\bibitem{Witten95} E. Witten, Nucl. Phys. {\bf B443}, 85 (1995). 

\bibitem{Tomita78} K. Tomita, Prog. Theor. Phys. {\bf 59}, 1150 (1978).

\bibitem{Veneziano97} G. Veneziano, Phys. Lett. {\bf B406}, 297 (1997),
hep-th/9703150.

\bibitem{MahOnoVen98} J. Maharana, E. Onofri, and G. Veneziano,
J. High Energy  Phys. {\bf 01}, 004 (1998), gr-qc/9802001.

\bibitem{BelKha69} V. A. Belinskii and I. M. Khalatnikov, Sov.
Phys. JETP {\bf  29}, 911 (1969).

\bibitem{BelKha70} V. A. Belinskii and I. M. Khalatnikov, Sov. Phys.
JETP {\bf 30}, 1174 (1970).

\bibitem{BelKha71} V. A. Belinskii and I. M. Khalatnikov, Sov. Phys.
JETP {\bf 32}, 169 (1971).

\bibitem{BelLifKha82} V. A. Belinskii, E. M. Lifshitz, and I. M.
Khalatnikov,
Adv. Phys. {\bf 31}, 639 (1982).

\bibitem{BarKun97a} J. D. Barrow and K. Kunze, Phys.
Rev. {\bf D56}, 741 (1997), hep-th/9701085.

\bibitem{FeiLazVaz97} A. Feinstein, R. Lazkoz, and M. A. Vazquez--Mozo,
Phys. Rev. {\bf D56}, 5166 (1997), hep-th/9704173.

\bibitem{ClaFeiLid99b} D. Clancy, A. Feinstein, J. E. Lidsey, and
R. Tavakol, Phys. Rev. {\bf D60}, 043503 (1999), gr-qc/9901062.

\bibitem{Bakas94} I. Bakas, Nucl. Phys. {\bf B428}, 374 (1994),
hep-th/9402016.

\bibitem{Maharana95} J. Maharana, Phys. Rev. Lett. {\bf 75}, 205 (1995),
hep-th/9502002.

\bibitem{Kehagias95} A. A. Kehagias, Phys. Lett. {\bf B360}, 19 (1995),
hep-th/9506205.

\bibitem{Geroch72} R. Geroch, J. Math. Phys. {\bf 13}, 394 (1972).

\bibitem{lidseyetal} J.E. Lidsey, D. Wands and E. Copeland, Phys. Rept. {\bf 337} (2000),
hep-th/9909061.

\bibitem{WaiIncMar79} J. Wainwright, W. C. W. Ince, and B. J.
Marshman, Gen. Rel. Grav. {\bf 10}, 259 (1979).

\bibitem{CarChaFei83} M. Carmeli, Ch. Charach, and A. Feinstein, 
Ann. Phys. {\bf 150}, 392 (1983). 

\bibitem{BCL} A. P. Billyard, A. A. Coley, and J. E. Lidsey, Phys. Rev. D. {\bf 59}, 123505  (1999);
{\em ibid}, J. Math. Phys. {\bf 40}, 5092 -5105 (1999); 
{\em ibid}, Class. Quant. Grav. {\bf 17}, 453-484 (2000).

\bibitem{MT} A. P. Billyard, A. A. Coley, J. E. Lidsey and U. S. Nilsson, Phys. Rev D 61, 043504 (2000).

\bibitem{FeinsteinIbanez93} A. Feinstein and J. Ibanez, Class. Quantum Grav., {\bf 10}, L227-L231, (1993)

\bibitem{kko} N. Kaloper, I. I. Kogan, and K. A. Olive, Phys. 
Rev. {\bf D57}, 7340 (1998); Erratum, ibid. {\bf D60}, 049901 (1999). 

\bibitem{Bthesis}A. P. Billyard, Ph. D thesis, Dalhousie University (1999)

\bibitem{coplahwan} E. J. Copeland, A. Lahiri, and D. Wands, 
Phys. Rev. {\bf D50}, 4868 (1994). 

\end{thebibliography}
\end{document}